\documentclass[aps,twocolumn,showpacs]{revtex4}
\makeatletter

\newcommand{\Rmnum}[1]{\expandafter\@slowromancap\romannumeral #1@}
\makeatother
\usepackage{graphicx}
\usepackage{dcolumn}
\usepackage{bm}
\usepackage{CJK}
\usepackage{amsfonts}
\usepackage{psfrag}
\usepackage{wrapfig}
\usepackage{subfigure}
\usepackage{makeidx}
\usepackage{multirow}
\usepackage{epsf}
\usepackage{hyperref}
\usepackage{amsmath}
\usepackage{cases}
\usepackage{multirow}

\begin{document}
\begin{CJK}{GBK}{song}

\newcommand{\eq}[1]{Eq.(\ref{#1})}
\newcommand{\ud}{\,\mathrm{d}\,}

\title{Breathers and solitons on two different backgrounds in a generalized coupled Hirota system with four wave mixing}

\author{Han-Xiang Xu$^{1,2}$, Zhan-Ying Yang$^{1,2}$\footnote{Corresponding author:zyyang@nwu.edu.cn}, Li-Chen Zhao$^{1,2}$, Liang Duan$^{1,2}$, Wen-Li Yang$^{2,3}$}

\address{School of Physics, Northwest University, Xi'an, 710127, China\\
Shaanxi Key Laboratory for Theoretical Physics Frontiers, Xi'an 710127, China \\
Institute of Modern Physics, Northwest University, Xi'an 710127, China}

\begin{abstract}
We study breathers and solitons on different backgrounds in optical fiber system, which is governed by generalized coupled Hirota equations with four wave mixing effect. On plane wave background, a transformation between different types of solitons is discovered. Then, on periodic wave background, we find breather-like nonlinear localized waves of which formation mechanism are related to the energy conversion between two components. The energy conversion results from four wave mixing. Furthermore, we prove that this energy conversion is controlled by amplitude and period of backgrounds. Finally, solitons on periodic wave background are also exhibited. These results would enrich our knowledge of nonlinear localized waves' excitation in coupled system with four wave mixing effect.

\end{abstract}


\maketitle

\section{Introduction}%
Hirota model plays a significant role in the study of nonlinear localized waves\cite{27,28,29}. Nowadays, many nonlinear localized waves governed by standard Hirota equation with novel dynamic features are observed in optical fiber system\cite{1,2,3}. Especially coupled Hirota equations, there are abundant nonlinear localized wave structures, such as solitons\cite{4,5,6,38,39}, breathers\cite{6}, rogue waves\cite{5,6,7,8,9,10} and composite structures\cite{6,7,8}. However, there are rare researches in Hirota system to consider a significant effect, four wave mixing (FWM) effect\cite{31,32}, which is studied in other nonlinear optical system\cite{30,11}. On the other hand, in Bose-Einstein condensate (BEC), the FWM term is paraphrased as a pair-transition term, which brings abundant nonlinear wave structures and dynamics on different backgrounds\cite{13,14,15}. Particularly some interesting phenomena emerge from the periodic background\cite{15}. Thus, it is important to study structures and dynamics of nonlinear localized wave in generalized coupled Hirota equations with FWM terms.

In this paper, we study several nonlinear localized wave structures on different backgrounds in generalized coupled Hirota system. On plane wave backgrounds, the transformation between different types of solitons is discovered. Moreover, on periodic wave backgrounds, the propagation of nonlinear localized waves exhibits varied temporal-spatial distributions and dynamic features. In particular, breather-like structures are observed on periodic backgrounds, which result from the energy conversion allowed by FWM in instable breather structures. However, even on periodic backgrounds there is no energy conversion in stable soliton structures. It should be noted that the interactions between nonlinear localized waves are not discussed in this paper, because some of interactions are similar with the interactions in other systems, which are reported by previous studies\cite{19,35}.

This manuscript will be structured as follows. In Sec. II, the generalized coupled Hirota equations with FWM terms and its analytical solution are given. In Sec. III, abundant nonlinear localized wave structures and dynamics, including transformation between different types of solitons, breather-like nonlinear localized waves and envelope-like solitons, are exhibited. And we unveil the formation mechanism of breather-like structures allowed by FWM. Finally, we summarize the results and present our conclusions in Sec. IV.

\section{The generalized coupled Hirota equations with four wave mixing terms and analytical solutions}%
Recently, the study of coupled nonlinear Schr\"odinger equations with a pair-transition term obtained from a standard nonlinear Schr\"odinger equation by a linear transformation has discovered some novel nonlinear localized wave structures\cite{12,13,14,15}. In this paper, we use the linear transformation on well known integrable standard Hirota equation to find out the integrable generalized coupled Hirota equations with FWM terms in optical fibers. This linear transformation and the integrability of coupled equations is demonstrated in Ref.\cite{17}. In dimensionless form, the generalized coupled Hirota equations with FWM terms read
\begin{eqnarray}
\begin{split}
&iu_{z}+\frac{1}{2}u_{tt}+(\left|{u}\right|^{2}+2\left|{v}\right|^{2})u+v^{2}u^{*} \\ &+i\beta[u_{ttt}+6(\left|{u}\right|^{2}+\left|{v}\right|^{2})u_{t}+6v^{*}v_{t}u+3(v^{2})_{t}u^{*}]=0, \label{equ:eq1}\
\end{split}
\\
\begin{split}
&iv_{z}+\frac{1}{2}v_{tt}+(2\left|{u}\right|^{2}+\left|{v}\right|^{2})v+u^{2}v^{*} \\ &+i\beta[v_{ttt}+6(\left|{u}\right|^{2}+\left|{v}\right|^{2})v_{t}+6u^{*}u_{t}v+3(u^{2})_{t}v^{*}]=0, \label{equ:eq2}\
\end{split}
\end{eqnarray}
where $u$ and $v$ are the complex envelopes of the two wave fields, $z$ is the propagation distance, $t$ is the retarded time and the symbol of $*$ represents the complex conjugation. The real parameter $\beta$ is introduced to be responsible for the high-order effect terms. If $\beta=0$, it reduces to the coupled nonlinear Schr\"odinger equations with a pair-transition term.

In other words, this linear transformation means that the Eqs. (\ref{equ:eq1}) and (\ref{equ:eq2}) can be decoupled into two standard Hirota equations of which form reads $iq_{z}+\frac{1}{2}(q_{tt}+2\left|{q}\right|^{2}q)+i\beta(q_{ttt}+6\left|{q}\right|^{2}q_{t})=0$. And the linear transformation is $u=\frac{1}{2}(q_{1}+q_{2})$ and $v=\frac{1}{2}(q_{1}-q_{2})$, where $q_{1}$ and $q_{2}$ are the analytical solutions of standard Hirota equations.

In previous studies\cite{14,15,16,17,18}, because of the linear transformation between coupled nonlinear equations and standard (scalar) nonlinear equation, the solutions of the coupled equations can be obtained from the solutions of standard equation by a linear superposition. Using this linear superposition, the analytical solutions of Eqs. (\ref{equ:eq1}) and (\ref{equ:eq2}) are expressed as
\begin{eqnarray}
\begin{split}
u=&\frac{Ae^{i\theta}+A'e^{i\theta'}}{2}-\bigg\{\Big[(4A^{2}+\alpha_{1}^{2}+\alpha_{2}^{2})\cos(\gamma) \\ &-4A\alpha_{1}\cosh(\xi)+i(4A^{2}-\alpha_{1}^{2}-\alpha_{2}^{2})\sin(\gamma) \\ &+4iA\alpha_{2}\sinh(\xi)\Big]\Big[4A\alpha_{1}\cos(\gamma) \\ &-(4A^{2}+\alpha_{1}^{2}+\alpha_{2}^{2})\cosh(\xi)\Big]^{-1}\bigg\}be^{i\theta}, \label{equ:eq3}\
\end{split}
\\
\begin{split}
v=&\frac{Ae^{i\theta}-A'e^{i\theta'}}{2}-\bigg\{\Big[(4A^{2}+\alpha_{1}^{2}+\alpha_{2}^{2})\cos(\gamma) \\ &-4A\alpha_{1}\cosh(\xi)+i(4A^{2}-\alpha_{1}^{2}-\alpha_{2}^{2})\sin(\gamma) \\ &+4iA\alpha_{2}\sinh(\xi)\Big]\Big[4A\alpha_{1}\cos(\gamma) \\ &-(4A^{2}+\alpha_{1}^{2}+\alpha_{2}^{2})\cosh(\xi)\Big]^{-1}\bigg\}be^{i\theta}, \label{equ:eq4}\
\end{split}
\end{eqnarray}
where
\begin{displaymath}
\begin{split}
&\xi=\zeta(t+\nu z),\quad \gamma=\sigma(t+\nu z),\\
&\nu=a-\frac{1}{2}\omega+\beta(4a^{2}-2A^{2}-4b^{2}-2a\omega+\omega^{2}),\\
&\zeta=\left(\frac{\sqrt{\mu^{2}+\chi^{2}}+\chi}{2}\right)^{\frac{1}{2}},\quad \sigma=\pm\left(\frac{\sqrt{\mu^{2}+\chi^{2}}-\chi}{2}\right)^{\frac{1}{2}},\\
&\chi=4b^{2}-4A^{2}-(2a+\omega)^{2},\quad\mu=-4b(2a+\omega),\\
&\alpha_{1}=2b+\zeta,\quad\alpha_{2}=\omega+2a+\sigma,\\
&\theta=kz+\omega t,\quad k=A^{2}-\frac{1}{2}\omega^{2}-6A^{2}\omega\beta+\omega^{3}\beta,\\
&\theta'=k'z+\omega' t,\quad k'=A'^{2}-\frac{1}{2}\omega'^{2}-6A'^{2}\omega'\beta+\omega'^{3}\beta,\\
&a=Re[\lambda],\quad b=Im[\lambda].
\end{split}
\end{displaymath}
\\$A, \omega$ and $k$ represent the amplitude, frequency and wave number of background electric field, respectively. $\lambda$ is the eigenvalue parameter in Darboux transformation. Eqs. (\ref{equ:eq3}) and (\ref{equ:eq4}) are formed by a nonlinear superposition of trigonometric functions ($\sin\gamma, \cos\gamma$) and hyperbolic functions ($\sinh\xi, \cosh\xi$) on the background of two plane waves ($Ae^{i\theta}, A'e^{i\theta'}$). Here the hyperbolic functions and trigonometric functions describe the localization and the periodicity of the nonlinear localized waves, respectively. For the two plane waves, if $A'=0$, Eqs. (\ref{equ:eq3}) and (\ref{equ:eq4}) become two same solutions that are the solution of standard Hirota equation multiplying $1/2$. So, excluding $A'e^{i\theta'}$, the remainder part of Eqs. (\ref{equ:eq3}) and (\ref{equ:eq4}) is called an initial nonlinear structure. Hence, $Ae^{i\theta}$ is called an initial plane wave and $A'e^{i\theta'}$ is called a later superposed plane wave. These analytical solutions contain abundant nonlinear localized wave structures, not only the classical nonlinear localized waves (anti-dark solitons, W-shaped solitons, multi-peak solitons, Akhmediev breathers, Kuznetsov-Ma breathers, Peregrine rogue waves, etc.) but also some coupled structures (breather-like nonlinear localized waves and envelope-like soliton) evolved from the classical one. In this paper, we only consider two typical classical structures\cite{26,33,34}, non-rational W-shaped soliton and anti-dark soliton, as an initial nonlinear structure. Additionally, non-rational W-shaped soliton requires $b>A$; anti-dark soliton requires $b<-A$ and both of them require $\omega=\frac{1}{6\beta}, a=-\frac{1}{2}\omega$. Obviously, if $\beta=0$, namely in coupled nonlinear Schr\"odinger equations, these nonlinear localized waves cannot exist. However, in this paper we are not discussing the high-order effect because we are aware of that, even though $\beta$ is assigned by different values, expect 0, the phenomena cannot be changed essentially.

\section{Several nonlinear localized waves on different backgrounds}

In this generalized coupled Hirota system, classical nonlinear localized wave structures still can be observed. On the other hand, some effects, which are similar with the pair-transition effect in Ref.\cite{13,14,15}, also appear in this generalized coupled Hirota system. However, we notice that when these classical nonlinear localized wave structures superposes on different backgrounds, especially periodic backgrounds, some interesting coupled structures of nonlinear localized waves are arising.
\begin{figure}[t]
\begin{center}
\subfigure{\includegraphics[width=0.45\textwidth]{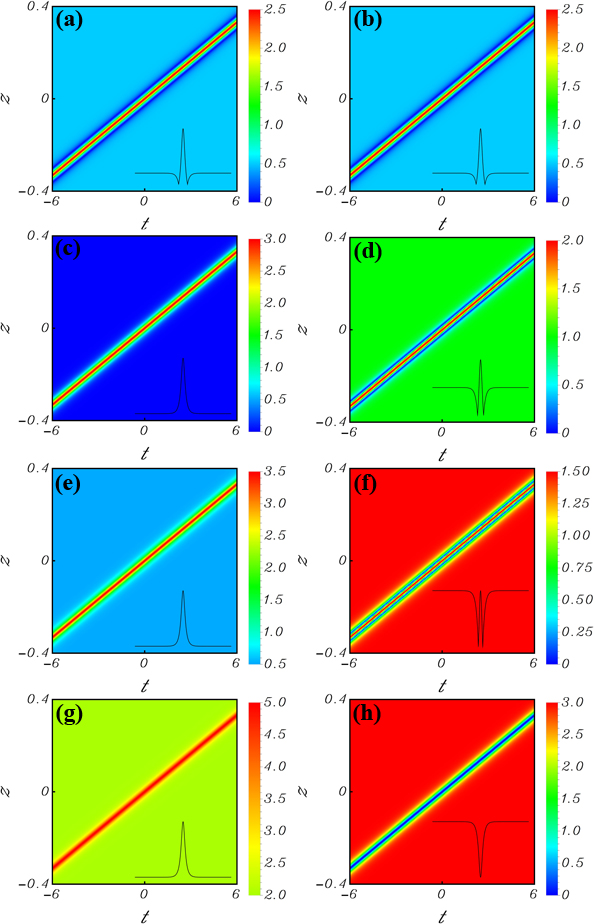}}
\caption{Transformation patterns of a non-rational W-shaped soliton on different superposed plane wave backgrounds, (a), (b) for $A'=0$, (c), (d) for $A'=A$, (e), (f) for $A'=\left|{b}\right|$, (g), (h) for $A'=\left|{A+2b}\right|$, and (a), (c), (e), (g) for component $u$, (b), (d), (f), (h) for component $v$.  In the lower right corner, there are cross sections of each patter along $t$ when $z=0$. It can be seen that non-rational W-shaped solitons transform to other soliton structures. The parameters are $\beta=1$, $\omega'=\omega=1/6$, $A=1$, $a=-1/12$, $b=2$.}
\label{fig:fig1}
\end{center}
\end{figure}

\subsection{Transformation between different types of solitons}

First of all, we study the dynamics of non-rational W-shaped soliton and anti-dark soliton on different plane wave backgrounds. It is required that the amplitudes of superposed plane waves ($A'$) can be changed and their frequencies are equal to the frequencies of initial plane waves ($\omega'=\omega$). By an analysis for Eqs. (\ref{equ:eq3}) and (\ref{equ:eq4}) in mathematics, it is evident that for the superposed plane waves, there are three special amplitude values, $A'=A$, $A'=\left|{b}\right|$ and $A'=\left|{A+2b}\right|$ (It determine the hump of initial soliton). These three values can be seen as critical values of different soliton structures in this coupled system. We give an example in Fig. \ref{fig:fig1}, which based on an initial structure of non-rational W-shaped soliton.

Depicting in Fig. \ref{fig:fig1}, there are different structures of components $u$ and $v$ when non-rational W-shaped soliton superposes on different plane wave backgrounds with increased amplitudes. It is easy to imagine that if $A'$ close to 0, components $u$ and $v$ will become two same non-rational W-shaped solitons (Fig. \ref{fig:fig1} (a), (b)). With increase of $A'$, component $u$ transforms to W-shaped soliton, bright soliton (Fig. \ref{fig:fig1} (c)) and anti-dark soliton (Fig. \ref{fig:fig1} (e), (g)) sequentially. These anti-dark solitons rise with background. And component $v$ transforms to W-shaped soliton (Fig. \ref{fig:fig1} (d), (f)) and dark soliton (Fig. \ref{fig:fig1} (h)) sequentially. It should be mentioned that the height of W-shaped soliton's hump are controlled by $A'$. When $A'=\left|{b}\right|$, the height of W-shaped soliton's hump equals to the background in component $v$. Besides, these complementary structures appearing in components $u$ and $v$ are derived from the coupled characteristic of this system. Namely the transformation between different types of solitons results from the four wave mixing effect. Entire structure transformation is classified in Table I.

In the meantime, based on an initial structure of anti-dark soliton, different transformation is described in Table II. Similarly, if $A'$ close to 0, components $u$ and $v$ will become two same anti-dark solitons. But with increase of $A'$, component $u$ transforms to diverse structures, as anti-dark soliton, bright soliton, W-shaped soliton and dark soliton sequentially. However component $v$ always remains as anti-dark soliton that rises with backgrounds. Importantly these basic structures in the both of two tables are related to the complex ones in the next sections.
\begin{table}[t]
\begin{center}
\caption{Transformation between different types of solitons based on non-rational W-shaped soliton. $A'$ is the amplitudes of superposed plane waves.}
\begin{tabular}{|c|c|c|}
\hline
\multirow{3}{0.15\textwidth}{\centering values of $A'$} & \multicolumn{2}{c|}{\multirow{2}{0.25\textwidth}{\centering Nonlinear localized waves \\type of each components}} \\ & \multicolumn{2}{c|}{} \\
\cline{2-3}
 & $u$ & $v$ \\
\hline
$[0,A)$ & W-shaped soliton & \multirow{3}{*}{\centering W-shaped soliton} \\
\cline{1-2}
$A$ & Bright soliton & \\
\cline{1-2}
$(A,\left|{A+2b}\right|)$ & \multirow{2}{*}{\centering Anti-dark soliton} & \\
\cline{1-1} \cline{3-3}
$[\left|{A+2b}\right|,+\infty)$ & & Dark soliton \\
\hline
\end{tabular}
\label{tab:tab1}
\end{center}
\end{table}
\begin{table}[t]
\begin{center}
\caption{Transformation between different types of solitons based on anti-dark soliton. $A'$ is the amplitudes of superposed plane waves.}
\begin{tabular}{|c|c|c|}
\hline
\multirow{3}{0.15\textwidth}{\centering values of $A'$} & \multicolumn{2}{c|}{\multirow{2}{0.25\textwidth}{\centering Nonlinear localized waves \\type of each components}} \\ & \multicolumn{2}{c|}{} \\
\cline{2-3}
 & $u$ & $v$ \\
\hline
$[0,A)$ & Anti-dark soliton & \multirow{4}{*}{\centering Anti-dark soliton} \\
\cline{1-2}
$A$ & Bright soliton & \\
\cline{1-2}
$(A,\left|{A+2b}\right|)$ & W-shaped soliton & \\
\cline{1-2}
$[\left|{A+2b}\right|,+\infty)$ & Dark soliton & \\
\hline
\end{tabular}
\label{tab:tab2}
\end{center}
\end{table}

\subsection{Breather-like patterns on periodic wave backgrounds}
\begin{figure*}[t]
\begin{center}
\subfigure{\includegraphics[width=0.75\textwidth]{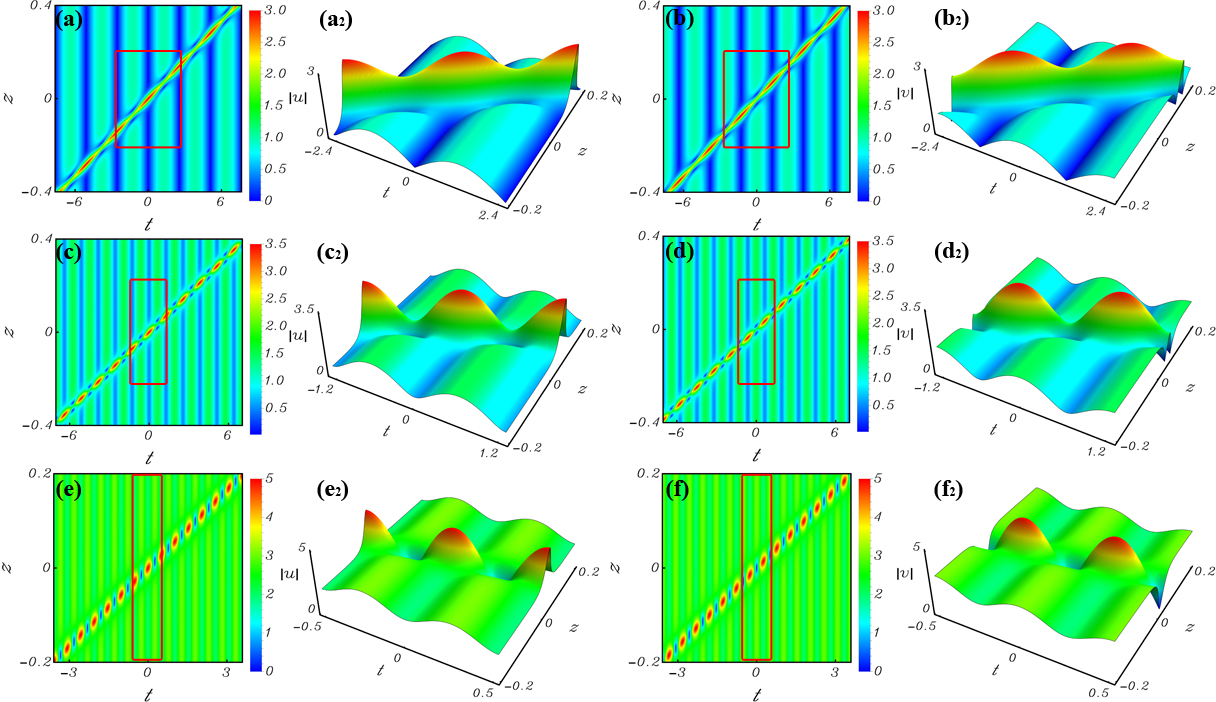}}
\caption{Breather-like patterns of a non-rational W-shaped soliton superposing on the periodic wave backgrounds along vertical direction, (a), (b) for $A'=A$, $\omega'=(1+\sqrt{219})/6$, (c), (d) for $A'=\left|{b}\right|$, $\omega'=(1+17\sqrt{3})/6$, (e), (f) for $A'=\left|{A+2b}\right|$, $\omega'=(1+\sqrt{5403})/6$, and (a), (c), (e) for component $u$, (b), (d), (f) for component $v$. (a$_{2}$), (b$_{2}$),(c$_{2}$),(d$_{2}$),(e$_{2}$) and (f$_{2}$) are three-dimensional patterns of red region of corresponding parts. Others parameters are $\beta=1$, $\omega=1/6$, $A=1$, $a=-1/12$, $b=2$.}
\label{fig:fig2}
\end{center}
\end{figure*}
\begin{figure*}[t]
\begin{center}
\subfigure{\includegraphics[width=0.75\textwidth]{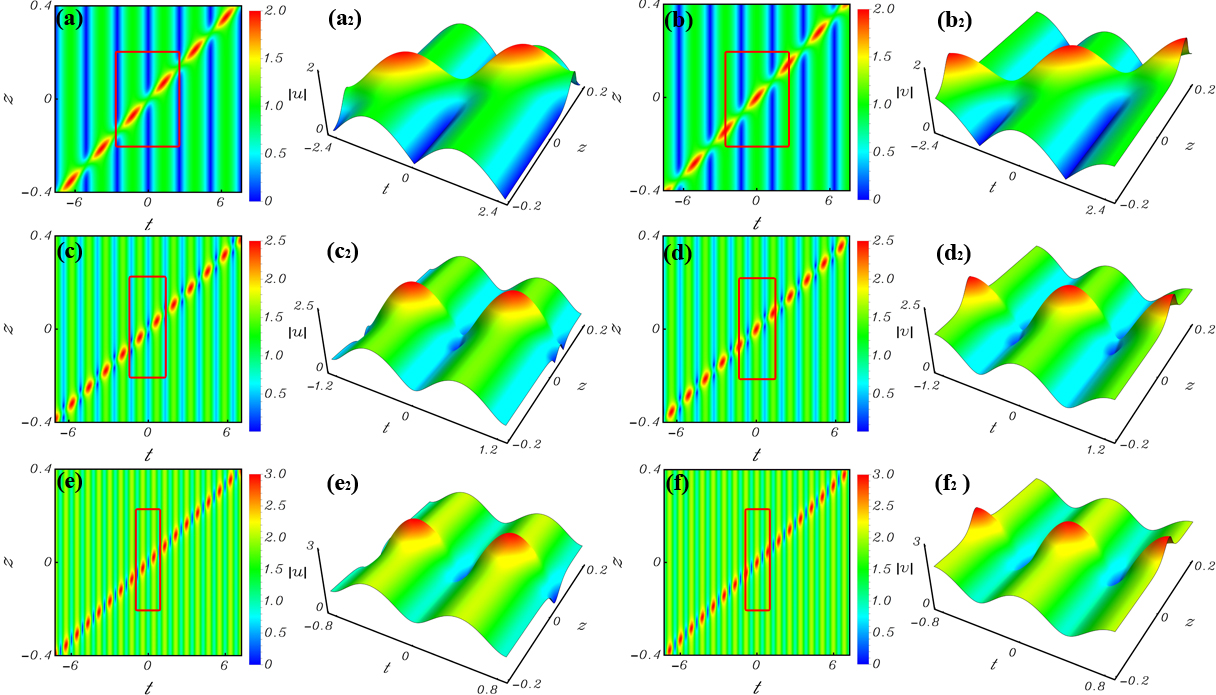}}
\caption{Breather-like patterns of an anti-dark soliton superposing on the periodic wave backgrounds along vertical direction, (a), (b) for $A'=A$, $\omega'=(1+\sqrt{219})/6$, (c), (d) for $A'=\left|{b}\right|$, $\omega'=(1+17\sqrt{3})/6$, (e), (f) for $A'=\left|{A+2b}\right|$, $\omega'=(1+\sqrt{1947})/6$, and (a), (c), (e) for component $u$, (b), (d), (f) for component $v$. (a$_{2}$), (b$_{2}$),(c$_{2}$),(d$_{2}$),(e$_{2}$) and (f$_{2}$) are three-dimensional patterns of red region of corresponding parts. Others parameters are same as Fig. 2, excluding $b=-2$.}
\label{fig:fig3}
\end{center}
\end{figure*}

In above section, we study solitons on plane wave backgrounds. Then, we consider the dynamics of non-rational W-shaped soliton and anti-dark soliton on different periodic wave backgrounds. In this section, the frequencies of superposed plane waves are unequal to the initial plane waves ($\omega'\not=\omega$), which results in the appearance of periodic wave backgrounds. For discussing conveniently, we select special values of $\omega'$ that makes periodic wave backgrounds along vertical direction. Fig. \ref{fig:fig2} is based on an initial structure of non-rational W-shaped soliton and Fig. \ref{fig:fig3} is based on anti-dark soliton.
\begin{figure}[t]
\begin{center}
\subfigure{\includegraphics[width=0.45\textwidth]{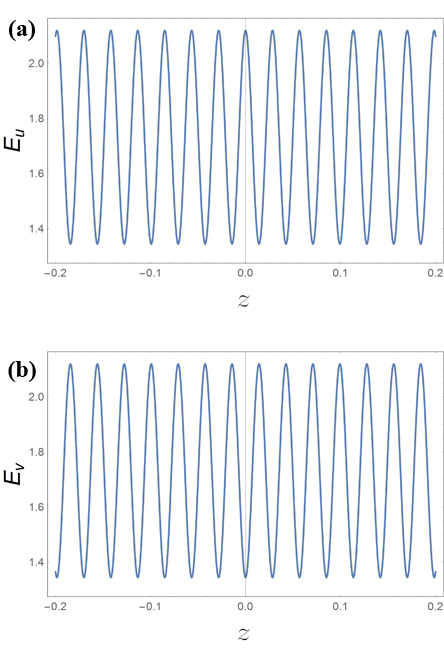}}
\caption{Energy exchanging patterns correspond to Fig. 3 (e) and (f) respectively, (a) for component $u$ and (b) for component $v$. It shows an energetic oscillation between components $u$ and $v$.}
\label{fig:fig4}
\end{center}
\end{figure}

In Fig. \ref{fig:fig2}, superposing and dynamic features of these interesting nonlinear localized wave structures will be discussed carefully. Firstly, periodic wave backgrounds are sine and cosine form for components $u$ and $v$ respectively. For example, along $t=0$, there is trough in component $u$ but crest in component $v$. So, two same nonlinear localized waves appear in components $u$ and $v$ with an only difference, a half-period movement between the backgrounds of two components, which is different from Fig. \ref{fig:fig1}. Secondly, kinds of breather-like nonlinear localized waves are observed in these periodic wave backgrounds, but they are not general breathers because their periods are not determined by themselves but controlled by the periods of backgrounds. These kinds of nonlinear localized waves can be seen as a non-rational W-shaped soliton superposed on periodic wave backgrounds. This superposition differs from previous studies about nonlinear localized waves superposing with periodic wave in uncoupled system or periodic modulation of nonlinear localized waves\cite{19,20,21,22,23}, because it is not a simple superposition or modulation. In two components, the breather-like nonlinear localized waves are controlled by backgrounds and interact with each other through backgrounds. Thirdly, we find some interesting structures in Fig. \ref{fig:fig2}. Remarkably, in Fig. \ref{fig:fig2}, peak-shape excitation appears in troughs but W-shape and valley-shape excitation emerge on crests. This phenomenon also comes from the superposition introduced by coupled system.

Similarly, based on an initial structure of anti-dark soliton, different breather-like nonlinear localized waves grow on periodic wave background. In Fig. \ref{fig:fig3}, when an anti-dark soliton superposed on periodic wave background, peak-shape excitation appears on crests but in troughs peak-shape, W-shape and valley-shape excitation are appearing sequentially.

Then we want to find the formation mechanism of these interesting breather-like nonlinear localized waves. Comparing the research in BEC where pair-transition effect allows particle conversion between two components \cite{13}, we infer that, in this case, four wave mixing effect allows energy conversion between components $u$ and $v$. Then, this energy conversion causes density distribution patterns in Fig. \ref{fig:fig2} and Fig. \ref{fig:fig3}. For proving this hypothesis, we give the perturbation energies of two components, which are defined as
\begin{eqnarray}
&E_{u}=\int_{-\infty}^{\infty}(\left|{u}\right|^{2}-\left|{u_{0}}\right|^{2})dt, \label{equ:eq5}\ \\
&E_{v}=\int_{-\infty}^{\infty}(\left|{v}\right|^{2}-\left|{v_{0}}\right|^{2})dt, \label{equ:eq6}\
\end{eqnarray}
where
\begin{flalign*}
&u_{0}=\frac{-Ae^{i\theta}+A'e^{i\theta'}}{2},\quad v_{0}=\frac{-Ae^{i\theta}-A'e^{i\theta'}}{2}.&
\end{flalign*}
$u_{0}$ and $v_{0}$ are backgrounds of each components. Using Eqs. (\ref{equ:eq5}) and (\ref{equ:eq6}), we give the energy exchanging patterns of Fig. \ref{fig:fig2} (e), (f) in Fig. \ref{fig:fig4} as an example. In Fig. \ref{fig:fig4}, we can find that perturbation energies of components $u$ and $v$ are complementary with each other. And their conversion periods perfectly correspond with the propagation period of breather-like nonlinear localized wave ($T=2\pi/\nu(\omega-\omega_{0})$). In other words, the propagation period represents the repeatability of breather-like nonlinear localized waves when they propagate on periodic wave background. This discovery also evidence that the periods of breather-like nonlinear localized waves are controlled by the periods of backgrounds. Then through calculating energy exchange of all patterns in Fig. \ref{fig:fig2} and Fig. \ref{fig:fig3}, we find three properties: ($\romannumeral1$) the midline of energy ($(E_{max}+E_{min})/2$) is a constant in all energy exchange patterns, which means total energy of two components is constant; ($\romannumeral2$) with increasing of $A'$, the quantity of energy exchange ($E_{max}-E_{min}$) is decreasing; ($\romannumeral3$) the quantity of energy exchange based on the initial structure of non-rational W-shaped soliton is larger than the anti-dark soliton. It can be easily understood that making a peak-shape excitation in the trough needs more energy than on the crest.
\begin{figure}[t]
\begin{center}
\subfigure{\includegraphics[width=0.45\textwidth]{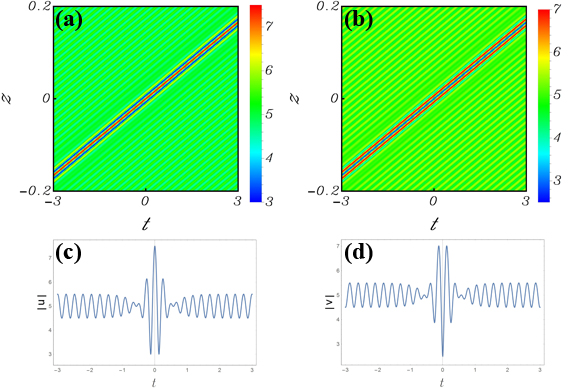}}
\caption{Envelope-like soliton patterns of a non-rational W-shaped soliton on the periodic wave backgrounds along the soliton, (a) for component $u$, (b) for component $v$ and (c), (d) are cross sections of (a), (b) along $t$ when $z=0$. The parameters are $\beta=1$, $\omega=1/6$, $\omega'= (1+6\sqrt{582})/6$, $A=1$, $A'=10$, $a=-1/12$, $b=2$.}
\label{fig:fig5}
\end{center}
\end{figure}
\begin{figure}[t]
\begin{center}
\subfigure{\includegraphics[width=0.45\textwidth]{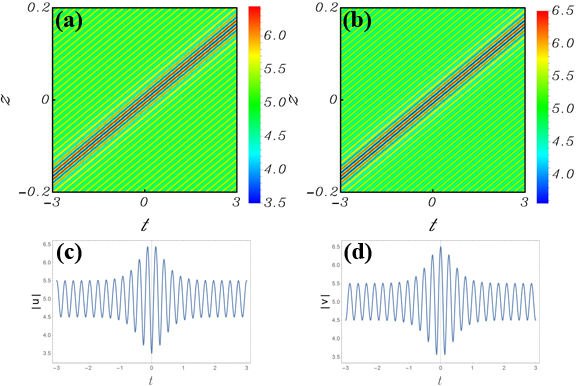}}
\caption{Envelope-like soliton patterns of an anti-dark soliton on the periodic wave backgrounds along the soliton, (a) for component $u$, (b) for component $v$ and (c), (d) are cross sections of (a), (b) along $t$ when $z=0$. The parameters are same as Fig. 5, excluding $b=-2$.}
\label{fig:fig6}
\end{center}
\end{figure}

In addition, we reveal a relationship between soliton structures on plane wave background and breather-like structures on periodic wave background. To explain the relationship, we choose Fig. \ref{fig:fig2} (a) and (b) as an example. We focus on limited temporal-spatial regions of overlapping parts between breath-like nonlinear localized wave and periodic wave background. Then we found that on crests, the profile and value of breath-like nonlinear localized wave are identical with the W-shaped soliton in Fig. \ref{fig:fig1} (d) and in troughs, the profile and value are identical with the bright soliton in Fig. \ref{fig:fig1} (c). For other patterns in Fig. \ref{fig:fig2} and Fig. \ref{fig:fig3}, the relationship is undifferentiated. On the crests of Fig. \ref{fig:fig2} (a), (c), (e) (or (b), (d), (f)), profiles and values of breath-like structures correspond with different types of solitons in Fig. \ref{fig:fig1} (d), (f), (h) respectively, i.e. the component $v$ in Table I. So, in the troughs of Fig. \ref{fig:fig2}, the profiles and values correspond with different types of solitons in component $u$ in Table I. Moreover, Fig. \ref{fig:fig3} corresponds with Table II. To summarize, when a soliton superpose with periodic wave backgrounds, a breather-like nonlinear localized wave emerges. And the breather-like nonlinear localized wave is related to the basic structures of this soliton on plane wave backgrounds. Especially on crests, the profiles of these breather-like nonlinear localized waves correspond with plane wave backgrounds' component $v$. And in troughs, the profiles correspond with plane wave backgrounds' component $u$.

\subsection{Envelope-like soliton patterns on periodic wave backgrounds}

Before clarifying superposition and dynamics of breath-like structures, we consider about some soliton structures on periodic wave backgrounds. These structures can be related to features of initial solitons when periodic wave backgrounds along the direction of solitons' velocities ($\nu$). To observe conveniently, we only show the situation that periodic wave backgrounds' wavelength is less than width of initial solitons. Based on non-rational W-shaped soliton is depicted in Fig. \ref{fig:fig5} and based on anti-dark soliton is depicted in Fig. \ref{fig:fig6}.

Some conclusions, which we have achieved before, can be used to explain these figures. Firstly, the half-period movement between components $u$ and $v$ is also remained. Secondly, in Fig. \ref{fig:fig5}, dark soliton is excited on crests and anti-dark soliton is excited in troughs corresponding to Table I. Meanwhile, in Fig. \ref{fig:fig6}, crests and troughs of periodic wave backgrounds are strengthened, which actually is an excitation of anti-dark soliton on crests and dark soliton in troughs corresponding to Table II. On the other hand, we calculate perturbation energies of two components, which, unlike Sec. III. B, are constant and no energy conversion because there is a stable soliton structure. The existence of conversion in stable or unstable structure is same to previous study\cite{13,24,35,36,37}. Besides, the profiles of superposition solitons discover a phenomenon that different effects on periodic wave backgrounds' profile depend on the different types of initial solitons. And the profile of solitons on periodic wave backgrounds are similar with the initial soliton's envelopes that can be found in Ref.\cite{25} Figure B1. So, these solitons on periodic wave backgrounds are called envelope-like solitons.

\section{Conclusions}

In conclusion, we have investigated the analytical solutions of generalized coupled Hirota equations with FWM terms, which are converted through a linear superposition from solutions of standard Hirota equation. On different backgrounds, especially periodic backgrounds, some interesting structures and dynamic features have been exhibited. We have presented the transformation between different types of solitons on plane wave backgrounds and shed light on the relationship between basic structures and complex breather-like structures. We have also revealed the formation mechanism of this superposition, which results from the energy conversion allowed by FWM. Finally, we have accounted that, when the directions of periodic backgrounds and solitons are undifferentiated, profiles of these solitons on periodic wave backgrounds are determined by the types of initial solutions. These findings may contribute to better understand structures and dynamics of nonlinear localized waves in the generalized coupled Hirota system with FWM and be extended to other coupled systems.

\section*{Acknowledgements}
This work is supported by the National Natural Science Foundation of China (NSFC)(Grant No. 11475135) and a special research project of the Education Department of Shaanxi Provincial Government (Contract No. 16JK1763).

\end{CJK}
\end{document}